\documentclass[usenatbib,useAMS]{mn2e}

\usepackage{graphicx,times}

\newcommand{\Hb}{\ensuremath{{\rm H}\beta}}
\newcommand{\Hg}{\ensuremath{{\rm H}\gamma}}
\newcommand{\HgA}{\ensuremath{{\rm H}\gamma_{\rm A}}}
\newcommand{\HgF}{\ensuremath{{\rm H}\gamma_{\rm F}}}
\newcommand{\Hd}{\ensuremath{{\rm H}\delta}}
\newcommand{\HdA}{\ensuremath{{\rm H}\delta_{\rm A}}}
\newcommand{\HdF}{\ensuremath{{\rm H}\delta_{\rm F}}}
\newcommand{\CNone}{\ensuremath{{\rm CN}_1}}
\newcommand{\CNtwo}{\ensuremath{{\rm CN}_2}}
\newcommand{\Mgtwo}{\ensuremath{{\rm Mg}_2}}
\newcommand{\Mgb}{\ensuremath{{\rm Mg}\, b}}
\newcommand{\TiOtwo}{\ensuremath{{\rm TiO}_2}}
\newcommand{\Fe}{\ensuremath{\langle {\rm Fe}\rangle}}
\newcommand{\aFe}{\ensuremath{\alpha/{\rm Fe}}}
\newcommand{\MgFep}{\ensuremath{[{\rm MgFe}]^{\prime}}}
\newcommand{\ZH}{\ensuremath{Z/{\rm H}}}

\title[Higher-order Balmer line indices in $\alpha$/Fe-enhanced stellar
population models] {\boldmath Higher-order Balmer line indices in
$\alpha$/Fe-enhanced stellar population models}

\author[Daniel Thomas, Claudia Maraston, \& Andreas Korn] {Daniel
Thomas, Claudia Maraston, \& Andreas Korn\\ Max-Planck-Institut f\"ur
extraterrestrische Physik, Giessenbachstra\ss e, D-85748 Garching,
Germany}

\date{Accepted 26 April 2004 Received ... ;
      in original form 27 November 2003}

\pagerange{\pageref{firstpage}--\pageref{lastpage}}

\pubyear{2004}

\begin{document}

\bibliographystyle{mn2e}
\maketitle

\label{firstpage}

\begin{abstract}
We have computed the higher-order Balmer absorption line indices \Hg\
and \Hd\ \citep{WO97} for stellar population models with variable
element ratios. The response of these indices to abundance ratio
variations is taken from detailed line formation and model atmosphere
calculations. We find that \Hg\ and \Hd, unlike \Hb, are very
sensitive to \aFe\ ratio changes at super-solar metallicities. Both
line indices increase significantly with increasing \aFe\ ratio. This
effect cannot be neglected when these line indices are used to derive
the ages of metal-rich, unresolved stellar populations like early-type
galaxies. We re-analyze the elliptical galaxy sample of \citet{Kun00},
and show that consistent age estimates from \Hb\ and \Hg\ are
obtained, only if the effect of \aFe\ enhancement on \Hg\ is taken
into account in the models. This result rectifies a problem currently
present in the literature, namely that \Hg\ and \Hd\ have up to now
led to significantly younger ages for early-type galaxies than
\Hb. Our work particularly impacts on the interpretation of
intermediate to high-redshift data, for which only the higher-order
Balmer lines are accessible observationally.
\end{abstract}


\section{Introduction}
\label{sec:intro}
Absorption line indices in the visual as defined by the Lick group
(e.g., \Hb, \Mgtwo, Fe5270, Fe5335, etc.,
\citealt{Buretal84,Fabetal85}) have proven to be a useful tool for the
derivation of ages and metallicities of unresolved stellar
populations. One of the largest merits of the Lick indices is to have
signaled the presence of non-solar Mg/Fe abundance ratios in the stars
of early-type galaxies
\citep*{WFG92,DSP93,CD94,FFI95,SB95,WPM95,Greggio97}.  This
interpretation is confirmed empirically by the similarity between the
Lick indices of early-type galaxies and those of Bulge globular
clusters \citep{Maretal03}, which are known from spectroscopy of
individual stars to be Mg/Fe enhanced.

Indeed the indices \Mgtwo/\Mgb\ and Fe5270/Fe5335 are dominated by
absorption lines from the elements Mg and Fe, and model atmosphere
calculations by \citet{Bar94} and \citet{TB95} have theoretically
demonstrated that these indices are sensitive to Mg/Fe abundance
ratios. Following an extension of the method introduced by
\citet{Traetal00a}, we have used the calculations of \citet{TB95} to
produce stellar population models with variable element abundance
ratios (\citealt*{TMB03a}, hereafter TMB03). These models are now able
to match simultaneously the Mg and Fe line indices of globular
clusters and early-type galaxies.

The Balmer line index \Hb, which is used as age indicator because it
measures the presence of warm A-type stars, is only moderately
contaminated by metallic lines and therefore relatively insensitive to
abundance ratio variations \citep[][TMB03]{TB95,Traetal00a}. Does this
convenient attribute of \Hb\ apply also to the higher-order Balmer
line indices \Hg\ and \Hd\ defined in \citet{WO97}?  The comparison of
the predictions from solar-scaled stellar population models with
galactic globular cluster data shows that the effect from
\aFe-enhancement must be small in a metallicity range up to solar
\citep{Maretal03}.  There are several good reasons for serious doubts,
however, that this is the case also at super-solar metallicities.  1)
From data in the literature (\citealt{KD98}; \citealt{Teretal99};
\citealt{Pogetal01a}; \citealt{Kunetal02a}; \citealt*{TFD04}) it can
be seen that the higher-order Balmer line indices and \Hb\ lead to
very inconsistent age estimates.  2) The \Hd\ measurements of
early-type galaxies in the Sloan Digital Sky Survey cannot be
reproduced by current solar-scaled stellar population models
\citep{Eisetal03}.  3) Both \Hg\ and \Hd\ have very prominent Fe
absorption features in their pseudo-continua.

\citet{TB95} have not included \Hg\ and \Hd, however, so that a
quantitative assessment of the sensitivity of these line indices to
abundance ratios in stellar population models was not possible until
now.  Since these line indices are already being widely used and will
certainly supersede \Hb\ as age indicators in future studies at
intermediate and high-redshifts, it is urgent to clarify this issue.
To this aim in Korn et al.\ (in preparation) we have extended the
\citet{TB95} approach and computed model atmosphere calculations with
variable abundance ratios including the wavelengths of the
higher-order Balmer lines ($\lambda\sim 4000\, {\rm \AA}$).  In this
paper we present the resulting stellar population models of \Hg\ and
\Hd\ with different \aFe\ ratios, which are computed following the
recipe of TMB03.

The paper is organised as follows. In Section~2 we will briefly
introduce the stellar population model, and calibrate it with galactic
globular clusters. In Section~3 we shall confront the new models with
data of elliptical galaxies focusing on the issues outlined above. We
will conclude with Section~4.

\section{The model}
TMB03 present stellar population models with different chemical
mixtures and element abundance ratios. All optical Lick indices from
\CNone\ to \TiOtwo\ \citep{Woretal94} in the wavelength range $4000\,
{\rm \AA}\la\lambda\la 6500\, {\rm \AA}$ are computed for various
chemical mixtures modifying the abundance ratios between
$\alpha$-elements (e.g., O, Mg, Si, Ti etc.), iron peak elements (Fe,
Cr), and the individual elements carbon, nitrogen, and calcium
\citep*{TMB03b}.  The models are based on the evolutionary population
synthesis of \citet{Ma98,Ma04}.  The impact from element ratio changes
is computed with the help of the \citet{TB95} model atmosphere
calculations, using an extension of the method introduced by
\citet{Traetal00a}.  We refer to TMB03 for more details.

\subsection{New model atmosphere calculations}
We have performed model atmosphere calculations based on the code
MAFAGS \citep{Gehren75a,Gehren75b} in a large range of metallicities
from 1/200 to 3.5 times solar, as presented and described in detail in
Korn et al.\ (in preparation).  In the work of \citet{TB95}, recently
defined absorption line indices like the higher-order Balmer lines
\citep{WO97} or the Ca$\,${\sc ii} triplet \citep{Cenetal01} are not
included.  Our new model atmosphere and line formation calculations
allow us to explore abundance ratio effects on any index
definition. Here we very briefly summarize the main results found for
the higher-order Balmer line indices \Hg\ and \Hd, relevant for this
paper.

We first calculated model atmospheres with solar-scaled element ratios
for various combinations of temperature, gravity, and total
metallicity. In subsequent models we double in turn the abundances of
the elements C, N, O, Mg, Fe, Ca, Na, Si, Cr, and Ti. We find that
higher-order Balmer line indices, unlike \Hb, are sensitive to element
abundance variations. More specifically, all four indices, \HgA, \HgF,
\HdA, and \HdF, show significant positive responses to the elements Mg
and Si, and significant negative responses to Fe, C, and Ti. The
latter are caused by the presence of absorption lines in the
pseudo-continua of the absorption index.  These sensitivities increase
with total metallicity because of the increasing strengths of the
metallic lines, and virtually disappear at low metallicities.

\subsection{Inclusion in the stellar population model}
\label{sec:newmodel}
In the TMB03 models, the \aFe-enhanced chemical mix at fixed total
metallicity is obtained by balancing the increased abundance of the
enhanced group with a decrease of the abundances of the elements Fe
and Cr (see also \citealt{Traetal00a}).  The former contains the
so-called $\alpha$-elements O, Ne, Mg, Si, S, Ar, Ca, Ti (particles
that are build up with $\alpha$-particle nuclei) plus the elements N
and Na.  The abundance of C (formally also an $\alpha$-particle) is
assumed not to vary (see TMB03 for more details). As the enhanced
group dominates by far the total metal abundance (mostly because O is
the most abundant metal), an enhancement of the \aFe\ ratio at fixed
total metallicity is characterized mainly by a decrease in the
abundances of Fe and Cr.  Hence, even though relative variations of
the element abundances within the enhanced group exist in early-type
galaxies \citep{Worthey98,Sagetal02,Sanetal03,TMB03b}, these do not
significantly affect the \aFe\ ratio.  This implies that the negative
responses of the higher-order Balmer line indices to Fe abundance as
reported above will dominate the behaviour of the indices as a
function of the \aFe\ ratio. A split of the abundance vectors into
further components exceeds the scope of this paper and will be subject
of future work.

The resulting stellar population models are shown in
Fig.~\ref{fig:globulars}, in which we plot the \aFe-independent index
\MgFep\ (see TMB03) vs.\ \HgA, \HgF, \HdA, and \HdF\ as functions of
total metallicity, for the abundance ratios $[\aFe]=0.0,\ 0.3,\ 0.5$
(see lables), and a fixed age of $12\,$Gyr. At high metallicities, the
models show a very prominent dependence of all four higher-order
Balmer line indices on the \aFe\ ratio.  This \aFe-sensitivity
disappears at low metallicities, because of the general weakness of Fe
and other metal absorption lines.

The effect at high metallicity amounts to the order of a couple of
Angstroms.  It is caused by the presence of a large number of iron
lines, the strongest being Fe$\,${\sc i} lines at $4308\,$\AA\ in the
blue and at 4383 and $4405\,$\AA\ in the red pseudo-continua.  The
most prominent Fe$\,${\sc i} lines in the index definitions of \HdA\
and \HdF\ are at 4045\, \AA\ (only \HdA), 4064 and 4072$\,$\AA\ in the
blue pseudo-continuum.  The red pseudo-continuum of the narrower index
\HdF\ contains Fe$\,${\sc i} lines at 4119 and 4132$\,$\AA, the red
pseudo-continuum of \HdA\ includes the lines at 4132 and 4144$\,$\AA.
\Hg\ seems generally slightly less affected, as this index has Fe and
Cr lines at 4326 and 4352 within the absorption feature, which may
counter-balance the effect from the iron lines in the continuum
windows.

In case of both \Hg\ and \Hd, the narrower index definitions are less
sensitive to abundance ratio variations.  It is obvious that the
narrower the definition of an index, the less it is affected by
contamination from additional metallic lines.  There is a price to
pay, however. Narrower indices require significantly higher
signal-to-noise ratios \citep{VA99}, which becomes very expensive in
terms of telescope time, and seriously hinders their use for galaxies
at intermediate and high redshifts.

\subsection{Check with globular clusters}
\begin{figure*}
\includegraphics[width=0.72\linewidth]{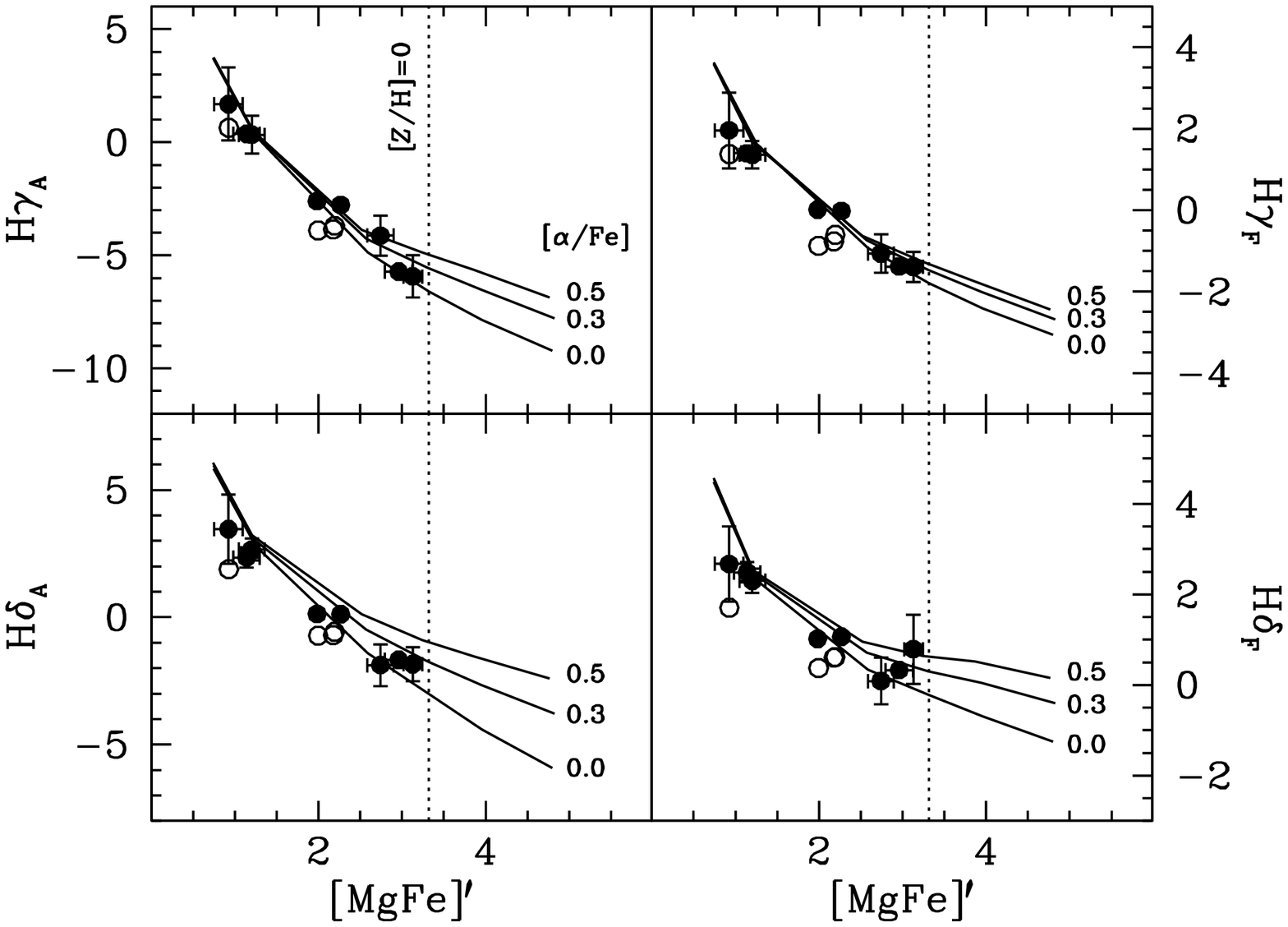}
\caption{The \aFe-independent index \MgFep\ vs.\ \HgA, \HgF, \HdA, and
\HdF\ as functions of total metallicity, for the abundance ratios
$[\aFe]=0.0,\ 0.3,\ 0.5$ (see lables), and a fixed age of $12\,$Gyr.
The vertical dotted line indicates the \MgFep\ of the solar
metallicity model.  Circles are the indices measured on the integrated
light of galactic globular clusters \citep{Puzetal02}. Open circles
are globular clusters with horizontal branches that are relatively red
given their metallicity \citep[see][]{Maretal03}. }
\label{fig:globulars}
\end{figure*}
The solar-scaled stellar population models of the higher-order Balmer
line indices are presented and compared with globular cluster data in
\citet{Maretal03}. Because of their weak sensitivity to \aFe\ ratios
at metallicities below solar (see Section~\ref{sec:newmodel}), the
solar-scaled models reproduce very well the data of globular clusters
up to solar metallicity (for the comparison at young ages see
\citealt*{BHS02}). Here we check whether also the new models, with the
effect of \aFe-enhancement included, are consistent with the globular
cluster data.  In Fig.~\ref{fig:globulars}, which is analogous to
Fig.~13 in \citet{Maretal03}, globular clusters \citep{Puzetal02} are
plotted as circles.  It is known from high-resolution spectroscopy of
individual stars in galactic globular clusters that these are
\aFe-enhanced by typically a factor 2 (see \citealt{Maretal03} and
references therein). The model can therefore be considered well
calibrated, at least for old ages, if the data are reproduced by the
model with $[\aFe]=0.3$.

The proper modelling of Balmer line indices requires the calibration
of the horizontal branch morphology (\citealt*{FB95,MT00,Leeetal00};
\citealt{Maretal03}).  In \citet{Maretal03} various morphologies of
the horizontal branch in the models are considered, that reproduce the
data with their observed horizontal branches. Since this issue is
fully discussed there, in Fig.~\ref{fig:globulars} we plot only the
model which best reproduces the average trend of the Balmer line
indices as a function of metallicity.  The open symbols are clusters
that have relatively red horizontal branches for their metallicity,
and it is shown in \citet{Maretal03} that these objects are well
matched by models in which less mass loss along the red giant branch
is adopted. Our model predictions should therefore be compared with
the filled data points. Hence, the models can be considered well
calibrated.

To conclude, as the sensitivity of \Hg\ and \Hd\ to \aFe\ disappears
at low metallicities and is still relatively small around solar
metallicity, both the solar-scaled and the \aFe-enhanced models are
consistent with the globular cluster data.

\section{The ages of elliptical galaxies}
\label{sec:ellipticals}
\begin{figure*}
\includegraphics[width=0.49\textwidth]{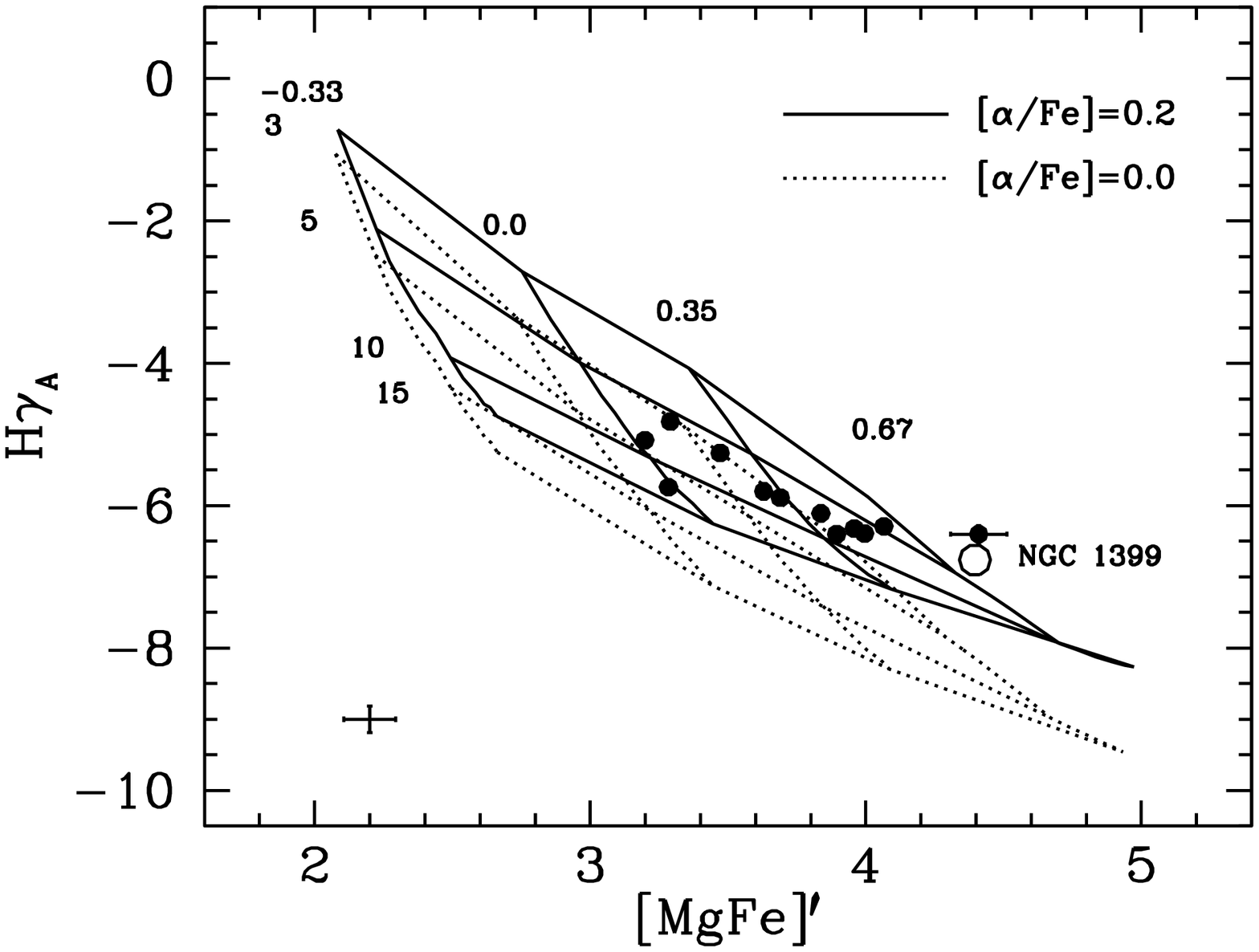}
\includegraphics[width=0.49\textwidth]{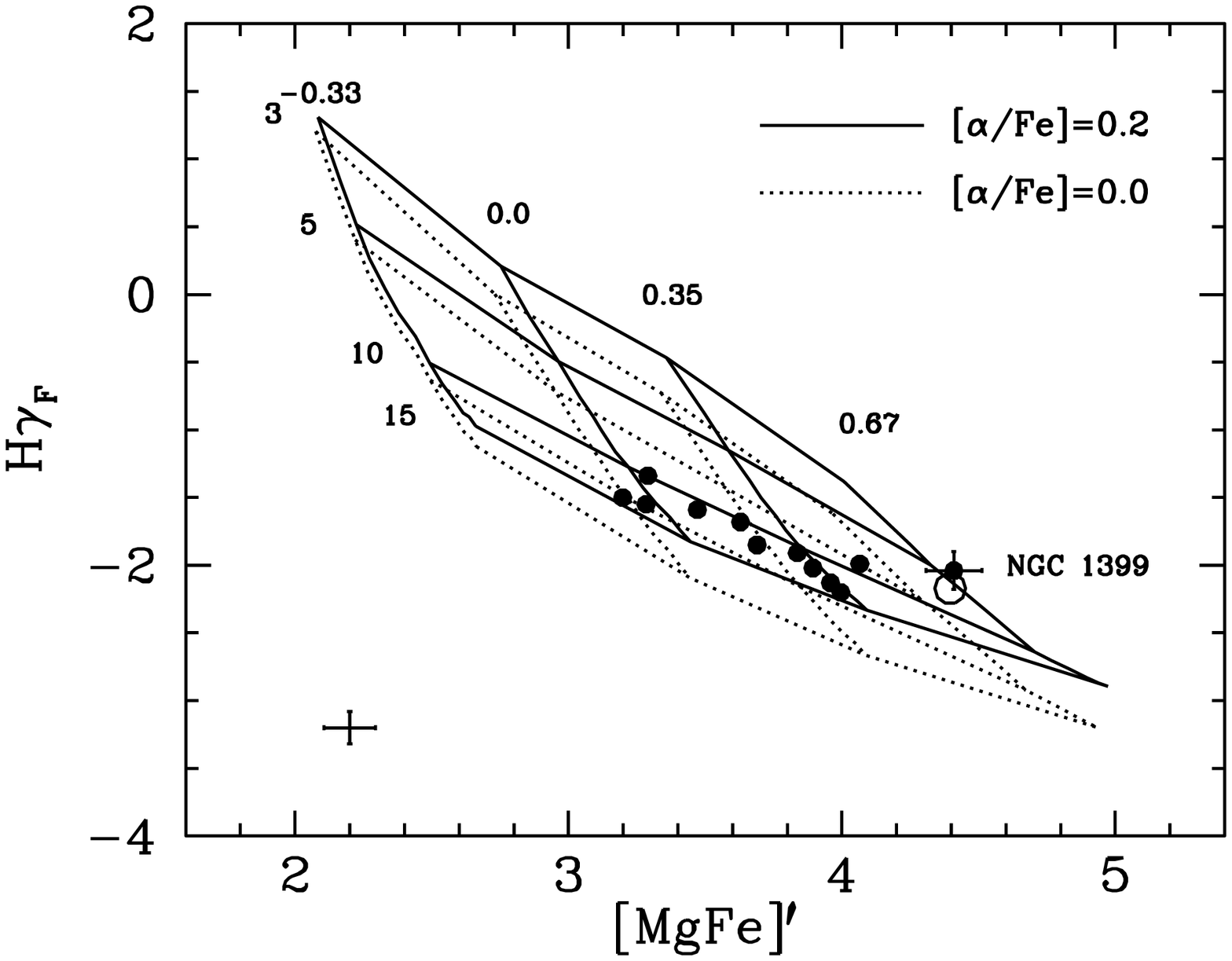}
\includegraphics[width=0.49\textwidth]{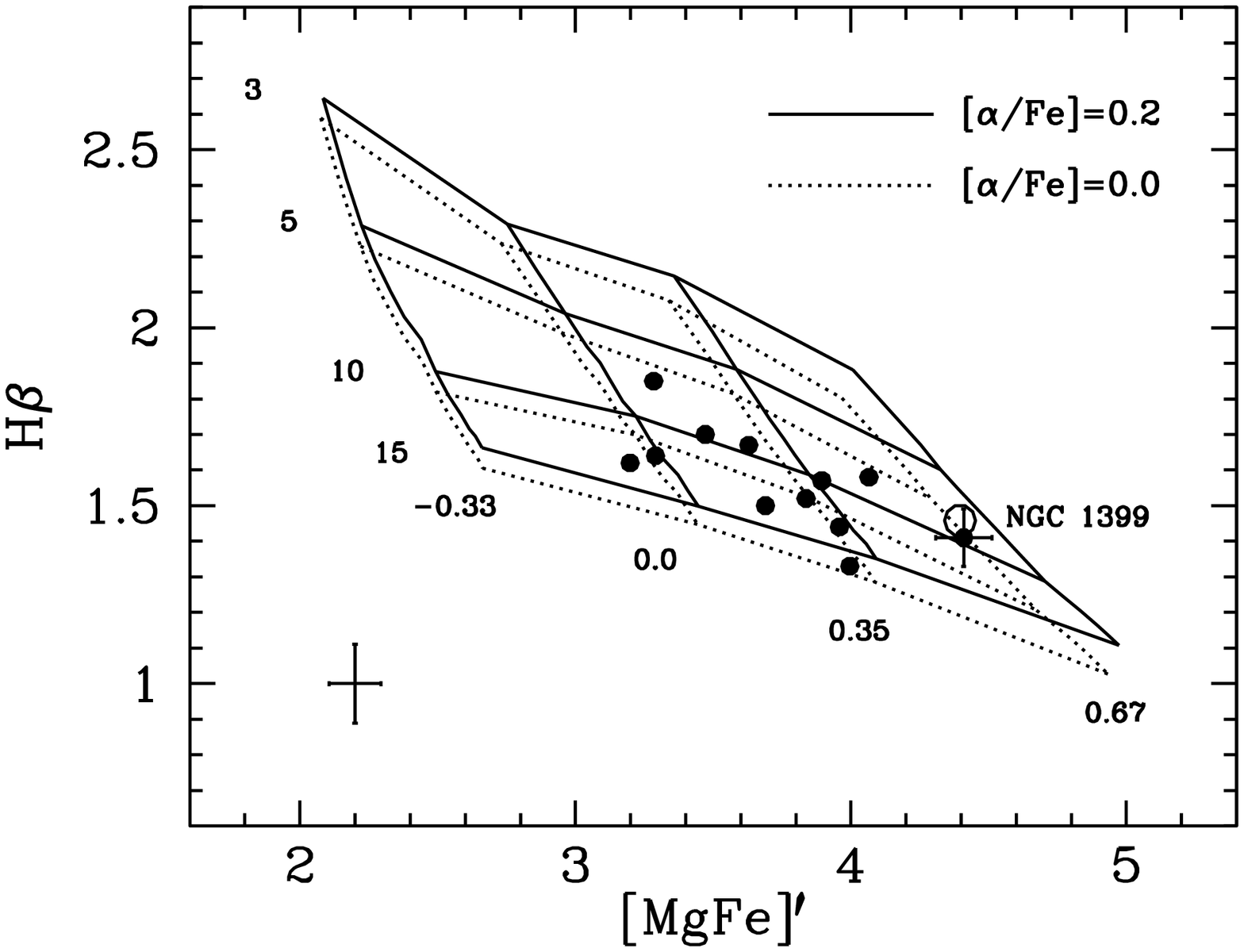}
\includegraphics[width=0.49\textwidth]{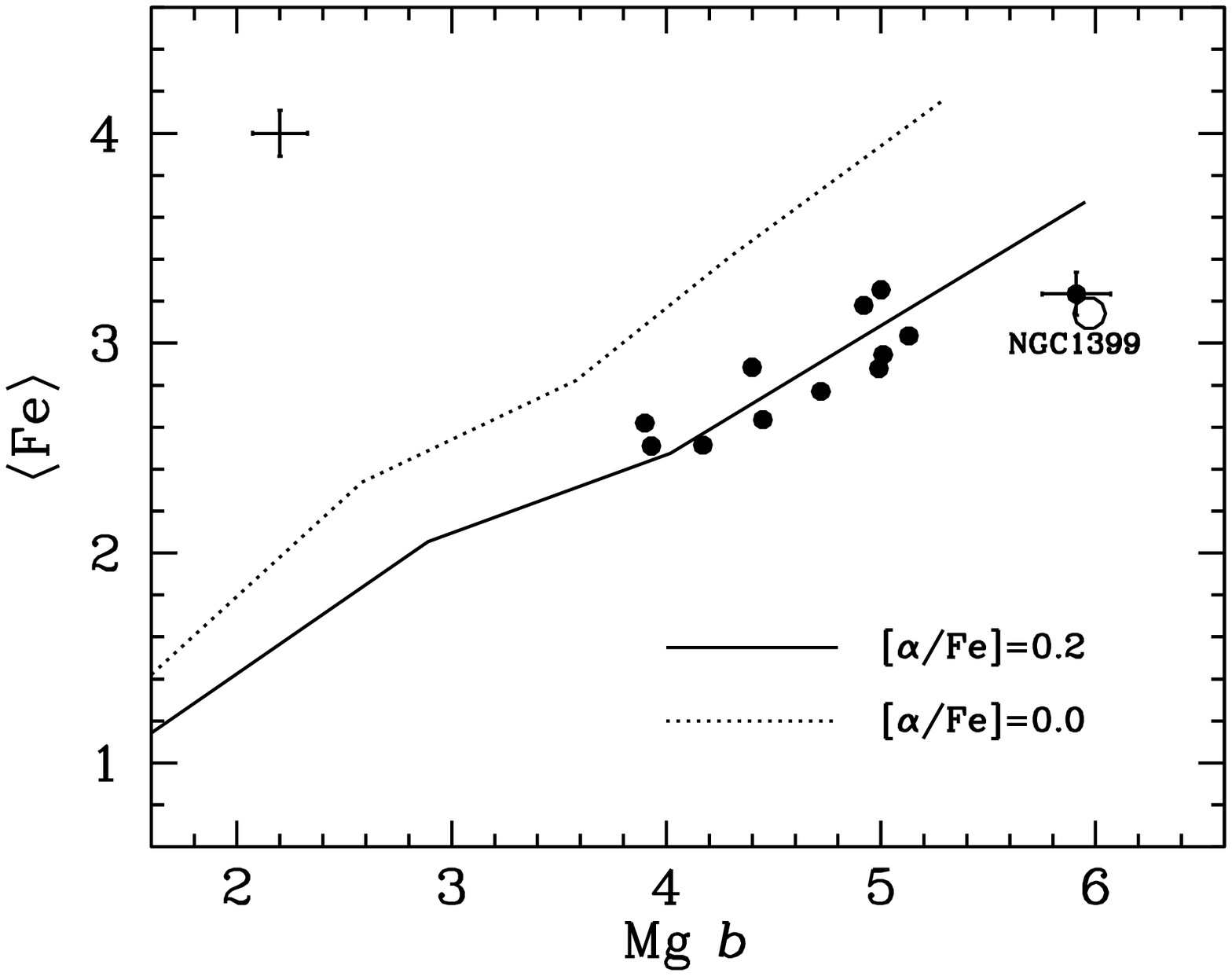}
\caption{Balmer line indices \HgA\ (top left-hand panel), \HgF\ (top
right-hand panel) and \Hb\ (bottom left-hand panel) as functions of
the \aFe-independent index \MgFep. The bottom right-hand panel shows
\Mgb\ vs.\ \Fe.  Solid and dotted lines are the \aFe-enhanced and the
solar-scales models, respectively. Models for the ages 3, 5, 10, and
15$\,$Gyr (only 10$\;$Gyr in the bottom right-hand panel), and the
metallicities $[\ZH]=-0.33,\ 0.0,\ 0.35,\ 0.67$ are shown (see
labels). Filled circles are elliptical galaxy data from
\citet{Kun00}. The open circle is a model with $t=9.5\,$Gyr,
$[\ZH]=0.58$ and $[\aFe]=0.38$ meant to reproduce NGC$\,$1399. The
error bars denote average 1-$\sigma$ errors. }
\label{fig:galaxies}
\end{figure*}
The effect of the \aFe\ ratio on the higher-order Balmer line indices
increases significantly with metallicity (Fig.~\ref{fig:globulars}),
so that a noticeable impact is to be expected on the derivation of
ages for metal-rich systems like elliptical galaxies. This is
illustrated in Fig.~\ref{fig:galaxies}, in which we plot the Balmer
line indices \HgA, \HgF, and \Hb\ of elliptical galaxies from the
Fornax cluster \citep{Kun00} versus their \MgFep\ indices. Overlayed
are solar-scaled (dotted lines) and \aFe-enhanced (solid lines) model
grids for ages from 3 to 15$\,$Gyr and total metallicities [\ZH]
between $-0.33$ and $0.67$ (see labels). It should be emphasized that
the elliptical galaxy data of \citet{Kun00} are particularly
appropriate for the purpose of this paper, as the sample displays an
amazingly small spread in \aFe\ ratio (see Fig.~\ref{fig:galaxies} and
text below). This allows us to show only one \aFe-enhanced model grid,
which considerably eases the illustration of the effect of the \aFe\
ratio on the age determination.

\subsection{\boldmath Based on solar-scaled models}
When interpreted with the solar-scaled models, the three Balmer line
indices yield dramatically discrepant age estimates: \HgA, \HgF, and
\Hb\ indicate average ages for the Fornax ellipticals of about 3, 8,
and 10$\;$Gyr, respectively.  In line with the different ages, also
inconsistent metallicities are obtained from the various Balmer lines,
\HgA\ leading to the highest values. NGC$\,$1399 (see label) is not
even covered by the solar-scaled models for \HgA\ and \HgF\, implying
metallicities well above $[\ZH]=0.67$ in extrapolation.

\subsection{\boldmath Based on \aFe-enhanced models}
However, from their Mg and Fe indices, we know that the \citet{Kun00}
ellipticals are all \aFe-enhanced with $[\aFe]\sim 0.2$ on the average
as shown in the bottom right-hand panel of Fig.~\ref{fig:galaxies}
(only NGC$\,$1399 has a slightly higher \aFe).  Hence, the appropriate
model to be used is the \aFe-enhanced model with $[\aFe]\sim 0.2$
shown as solid lines in Fig.~\ref{fig:galaxies}. These models predict
substantially larger (up to 1$\;$\AA) \HgA, still significantly larger
\HgF, and only moderately larger \Hb\ indices than the solar-scaled
ones. As a consequence, the severe age discrepancies obtained on the
basis of the solar-scaled model are eliminated. When interpreted with
the \aFe-enhanced model, the age and metallicity estimates from \Hb\
and \HgF\ now agree amazingly well. We find the ages to scatter
between 10 and 15$\;$Gyr with metallicities between one and two times
solar. The same good agreement holds also for NGC$\,$1399.  To
demonstrate this, we plot as the open circle in
Fig.~\ref{fig:galaxies} the model prediction with the three parameters
$t=9.5\,$Gyr, $[\ZH]=0.58$ and $[\aFe]=0.38$, which reproduces well
(within the errors) all six absorption line indices.

The agreement with \HgA\ is less compelling, even though the situation
has clearly improved with respect to the solar-scaled model. The age
estimates are roughly consistent within the errors, but we still
obtain systematically younger ages from \HgA\ than from \Hb\ and \HgF.
This might be due to an underestimation of the index responses drawn
from our model atmosphere calculations (Korn et al., in
preparation). However, it cannot be excluded that the fitting
functions for \HgA\ are in error at metallicities above solar owing to
bad input stellar data, or that calibration offsets are present in the
galaxy data.  The former caution can expect clarification, when in
future index response functions of the higher-order Balmer lines from
other groups will be available \citep{Houetal02,TFD04}.  To conclude,
at this point it is certainly safer to use the narrower definition
\HgF, if the cost in signal-to-noise is bearable.

\subsection{Age determinations in the literature}
The content of this paper strongly impacts on some results reported in
the literature, which have been obtained by using higher-order Balmer
line indices as age indicators. In the previous sections, we have
extensively discussed the interpretation of the \citet{Kun00}
sample. A further example is the work by \citet{Teretal99}, who
analyze the \HgA\ and \HdA\ indices of a sample of $\sim 100$ Coma
galaxies. Interpreting the data with solar-scaled stellar population
models, they find relatively young ages for early-type cluster
galaxies between 3 and 5$\,$Gyr, and in particular they find the trend
that more luminous objects tend to be younger. However, a well-defined
correlation between \aFe-enhancement and galaxy mass is found for
early-type galaxies (\citealt{Traetal00b}; \citealt*{TMB02b};
\citealt{Mehetal03}). Hence, both the young ages and the
anti-correlation between age and galaxy luminosity disappear, when the
effect from \aFe\ ratios on the higher-order Balmer line indices is
taken into account.

As a further example, \citet{Eisetal03} have recently measured
absorption line indices of early-type spectra from the Sloan Digital
Sky Survey. They find severe inconsistencies between several line
indices. Both \Hb\ and \Mgb\ cannot be matched simultaneously with the
higher-order Balmer line index \HdA\ by solar-scaled stellar
population models for any age-metallicity combination.  This problem
is caused, at least partially, by abundance ratios effects and can be
resolved with the models presented in this paper. We have shown that
\HdA\ increases by several Angstroms when the \aFe\ ratio is doubled.
Note that a more quantitative comparison is not possible, because the
data presented in \citet{Eisetal03} are not strictly calibrated on the
Lick system.

It has also been noticed in the literature that early-type galaxy data
are better matched by the solar-scaled models, when higher-order
Balmer line indices are plotted as functions of Fe indices (Fe4383,
Fe5270) instead \citep[e.g.,][]{Kun00,Pogetal01a,Eisetal03}.  From the
models presented here we know now that this apparent consistency
between data and models is merely the consequence of two compensating
effects: The increase of the higher-order Balmer line indices owing to
the \aFe\ enhancement is balanced by a decrease of the Fe line
indices, which conspires to locate the data on the model grid.

\section{Conclusions}
We have extended our models of absorption line indices of stellar
populations with variable abundance ratios (TMB03) to the higher-order
Balmer line indices \HgA, \HgF, \HdA, and \HdF. We show that the model
predictions are well consistent with data of galactic globular
clusters.  The key result of this work is that, unlike \Hb, all four
indices show a marked dependence on the \aFe\ ratio. The significance
of this effect increases with increasing metallicity, and therefore
impacts on the age derivation for elliptical galaxies.  Re-analyzing
the elliptical galaxy sample of \citet{Kun00}, we show that \Hb\ and
\HgF\ (and a bit less convincingly also \HgA) yield consistent age
estimates if (and only if) the effect of \aFe\ enhancement is taken
into account.

\medskip
A posteriori, it is not surprising that blue absorption line indices
are more affected by abundance ratio effects, because of the presence
of more metallic lines in the bluer parts of the spectrum. This fact
certainly diminishes the advantage of the higher-order Balmer line
indices over \Hb\ as age indicators, in spite of their lower
sensitivity to emission line filling.  The consequence for the age
derivation by means of \Hg\ and \Hd\ is twofold: 1) The element ratio
effect on the Balmer line indices needs to be taken into account in
the models.  2) The additional knowledge of the \aFe\ ratio is
necessary, which requires the simultaneous measurement of \Mgb\ and
Fe5270/Fe5335. If these wavelengths are not accessible, as typically
the case for intermediate and high-redshift data, \aFe\ ratios may be
best estimated from the indices Fe4383 and \CNone\ or \CNtwo.

\medskip
Model tables are listed in the Appendix and are available
electronically under http://www.mpe.mpg.de/$\sim$dthomas.

\section*{Acknowledgements}
We are grateful to the referee, Guy Worthey, for his very constructive
report and to Scott Trager for the continuously stimulating
discussions.

\bsp
\label{lastpage}

\appendix
\section{Model tables}
\begin{table*}
\caption{Simple stellar population models for the higher-order Balmer
lines with $t=1\;$--$\;15\;$Gyr, $[\ZH]=-2.25\;$--$\;0.67$, and
$[\aFe]=0$.}
\begin{minipage}{0.49\textwidth}
\begin{tabular}{rrrrrrr}
\hline\hline
\multicolumn{1}{c}{Age} & \multicolumn{1}{c}{[\ZH]} & \multicolumn{1}{c}{[\aFe]} & \multicolumn{1}{c}{\HdA} & \multicolumn{1}{c}{\HdF} & \multicolumn{1}{c}{\HgA} & \multicolumn{1}{c}{\HgF}\\
\multicolumn{1}{c}{(1)} & \multicolumn{1}{c}{(2)}   & \multicolumn{1}{c}{(3)}    & \multicolumn{1}{c}{(4)}  & \multicolumn{1}{c}{(5)}  & \multicolumn{1}{c}{(6)} & \multicolumn{1}{c}{(7)}\\
\hline
$ 1$ & $-2.25$ & $ 0.00$ & $ 10.53$ & $  8.26$ & $ 10.48$ & $  8.11$\\
$ 2$ & $-2.25$ & $ 0.00$ & $  8.38$ & $  6.29$ & $  7.70$ & $  6.06$\\
$ 3$ & $-2.25$ & $ 0.00$ & $  8.07$ & $  5.90$ & $  6.88$ & $  5.50$\\
$ 4$ & $-2.25$ & $ 0.00$ & $  7.65$ & $  5.53$ & $  6.25$ & $  5.09$\\
$ 5$ & $-2.25$ & $ 0.00$ & $  7.12$ & $  5.14$ & $  5.55$ & $  4.65$\\
$ 6$ & $-2.25$ & $ 0.00$ & $  6.60$ & $  4.77$ & $  4.86$ & $  4.24$\\
$ 7$ & $-2.25$ & $ 0.00$ & $  6.06$ & $  4.44$ & $  4.30$ & $  3.90$\\
$ 8$ & $-2.25$ & $ 0.00$ & $  5.71$ & $  4.23$ & $  4.00$ & $  3.70$\\
$ 9$ & $-2.25$ & $ 0.00$ & $  5.40$ & $  4.19$ & $  3.35$ & $  3.45$\\
$10$ & $-2.25$ & $ 0.00$ & $  5.57$ & $  4.25$ & $  3.33$ & $  3.43$\\
$11$ & $-2.25$ & $ 0.00$ & $  5.77$ & $  4.38$ & $  3.63$ & $  3.56$\\
$12$ & $-2.25$ & $ 0.00$ & $  5.83$ & $  4.48$ & $  3.75$ & $  3.62$\\
$13$ & $-2.25$ & $ 0.00$ & $  5.61$ & $  4.35$ & $  3.60$ & $  3.48$\\
$14$ & $-2.25$ & $ 0.00$ & $  5.20$ & $  4.10$ & $  3.07$ & $  3.15$\\
$15$ & $-2.25$ & $ 0.00$ & $  4.97$ & $  3.94$ & $  2.86$ & $  2.98$\\
$ 1$ & $-1.35$ & $ 0.00$ & $  7.83$ & $  5.84$ & $  6.98$ & $  5.67$\\
$ 2$ & $-1.35$ & $ 0.00$ & $  7.44$ & $  5.41$ & $  6.40$ & $  5.24$\\
$ 3$ & $-1.35$ & $ 0.00$ & $  6.72$ & $  4.89$ & $  5.36$ & $  4.64$\\
$ 4$ & $-1.35$ & $ 0.00$ & $  5.79$ & $  4.29$ & $  4.15$ & $  3.96$\\
$ 5$ & $-1.35$ & $ 0.00$ & $  4.88$ & $  3.73$ & $  2.94$ & $  3.29$\\
$ 6$ & $-1.35$ & $ 0.00$ & $  4.19$ & $  3.30$ & $  2.05$ & $  2.78$\\
$ 7$ & $-1.35$ & $ 0.00$ & $  3.62$ & $  2.95$ & $  1.33$ & $  2.37$\\
$ 8$ & $-1.35$ & $ 0.00$ & $  3.29$ & $  2.75$ & $  0.93$ & $  2.13$\\
$ 9$ & $-1.35$ & $ 0.00$ & $  2.93$ & $  2.54$ & $  0.42$ & $  1.84$\\
$10$ & $-1.35$ & $ 0.00$ & $  2.75$ & $  2.40$ & $  0.17$ & $  1.68$\\
$11$ & $-1.35$ & $ 0.00$ & $  2.70$ & $  2.34$ & $  0.10$ & $  1.62$\\
$12$ & $-1.35$ & $ 0.00$ & $  2.79$ & $  2.32$ & $  0.27$ & $  1.63$\\
$13$ & $-1.35$ & $ 0.00$ & $  3.06$ & $  2.54$ & $  0.57$ & $  1.83$\\
$14$ & $-1.35$ & $ 0.00$ & $  3.76$ & $  3.01$ & $  1.36$ & $  2.27$\\
$15$ & $-1.35$ & $ 0.00$ & $  4.08$ & $  3.22$ & $  1.70$ & $  2.44$\\
$ 1$ & $-0.33$ & $ 0.00$ & $  5.52$ & $  4.11$ & $  3.95$ & $  3.84$\\
$ 2$ & $-0.33$ & $ 0.00$ & $  2.62$ & $  2.52$ & $  0.35$ & $  1.95$\\
$ 3$ & $-0.33$ & $ 0.00$ & $  1.47$ & $  1.90$ & $ -1.06$ & $  1.20$\\
$ 4$ & $-0.33$ & $ 0.00$ & $  0.79$ & $  1.50$ & $ -1.92$ & $  0.71$\\
$ 5$ & $-0.33$ & $ 0.00$ & $  0.40$ & $  1.28$ & $ -2.48$ & $  0.40$\\
$ 6$ & $-0.33$ & $ 0.00$ & $  0.08$ & $  1.10$ & $ -2.95$ & $  0.15$\\
$ 7$ & $-0.33$ & $ 0.00$ & $ -0.23$ & $  0.94$ & $ -3.36$ & $ -0.08$\\
$ 8$ & $-0.33$ & $ 0.00$ & $ -0.46$ & $  0.82$ & $ -3.67$ & $ -0.25$\\
$ 9$ & $-0.33$ & $ 0.00$ & $ -0.70$ & $  0.70$ & $ -3.99$ & $ -0.43$\\
$10$ & $-0.33$ & $ 0.00$ & $ -0.99$ & $  0.55$ & $ -4.35$ & $ -0.63$\\
$11$ & $-0.33$ & $ 0.00$ & $ -1.23$ & $  0.44$ & $ -4.65$ & $ -0.79$\\
$12$ & $-0.33$ & $ 0.00$ & $ -1.42$ & $  0.34$ & $ -4.87$ & $ -0.93$\\
$13$ & $-0.33$ & $ 0.00$ & $ -1.57$ & $  0.28$ & $ -5.05$ & $ -1.02$\\
$14$ & $-0.33$ & $ 0.00$ & $ -1.63$ & $  0.27$ & $ -5.11$ & $ -1.05$\\
$15$ & $-0.33$ & $ 0.00$ & $ -1.74$ & $  0.23$ & $ -5.25$ & $ -1.12$\\
\hline
\end{tabular}
\end{minipage}
\begin{minipage}{0.49\textwidth}
\begin{tabular}{rrrrrrr}
\hline\hline
\multicolumn{1}{c}{Age} & \multicolumn{1}{c}{[\ZH]} & \multicolumn{1}{c}{[\aFe]} & \multicolumn{1}{c}{\HdA} & \multicolumn{1}{c}{\HdF} & \multicolumn{1}{c}{\HgA} & \multicolumn{1}{c}{\HgF}\\
\multicolumn{1}{c}{(1)} & \multicolumn{1}{c}{(2)}   & \multicolumn{1}{c}{(3)}    & \multicolumn{1}{c}{(4)}  & \multicolumn{1}{c}{(5)}  & \multicolumn{1}{c}{(6)} & \multicolumn{1}{c}{(7)}\\
\hline
$ 1$ & $ 0.00$ & $ 0.00$ & $  5.54$ & $  4.01$ & $  3.88$ & $  3.74$\\
$ 2$ & $ 0.00$ & $ 0.00$ & $  1.09$ & $  1.72$ & $ -1.53$ & $  0.95$\\
$ 3$ & $ 0.00$ & $ 0.00$ & $ -0.31$ & $  1.02$ & $ -3.35$ & $  0.00$\\
$ 4$ & $ 0.00$ & $ 0.00$ & $ -0.81$ & $  0.76$ & $ -4.04$ & $ -0.37$\\
$ 5$ & $ 0.00$ & $ 0.00$ & $ -1.29$ & $  0.53$ & $ -4.67$ & $ -0.71$\\
$ 6$ & $ 0.00$ & $ 0.00$ & $ -1.70$ & $  0.34$ & $ -5.16$ & $ -0.98$\\
$ 7$ & $ 0.00$ & $ 0.00$ & $ -1.91$ & $  0.24$ & $ -5.41$ & $ -1.12$\\
$ 8$ & $ 0.00$ & $ 0.00$ & $ -2.06$ & $  0.17$ & $ -5.58$ & $ -1.23$\\
$ 9$ & $ 0.00$ & $ 0.00$ & $ -2.32$ & $  0.06$ & $ -5.88$ & $ -1.39$\\
$10$ & $ 0.00$ & $ 0.00$ & $ -2.51$ & $ -0.02$ & $ -6.08$ & $ -1.51$\\
$11$ & $ 0.00$ & $ 0.00$ & $ -2.78$ & $ -0.13$ & $ -6.37$ & $ -1.67$\\
$12$ & $ 0.00$ & $ 0.00$ & $ -3.01$ & $ -0.23$ & $ -6.59$ & $ -1.79$\\
$13$ & $ 0.00$ & $ 0.00$ & $ -3.21$ & $ -0.31$ & $ -6.78$ & $ -1.90$\\
$14$ & $ 0.00$ & $ 0.00$ & $ -3.40$ & $ -0.39$ & $ -6.96$ & $ -2.00$\\
$15$ & $ 0.00$ & $ 0.00$ & $ -3.58$ & $ -0.46$ & $ -7.12$ & $ -2.09$\\
$ 1$ & $ 0.35$ & $ 0.00$ & $  2.87$ & $  2.55$ & $  0.40$ & $  1.91$\\
$ 2$ & $ 0.35$ & $ 0.00$ & $ -0.90$ & $  0.89$ & $ -3.93$ & $ -0.21$\\
$ 3$ & $ 0.35$ & $ 0.00$ & $ -1.62$ & $  0.52$ & $ -4.87$ & $ -0.72$\\
$ 4$ & $ 0.35$ & $ 0.00$ & $ -2.24$ & $  0.24$ & $ -5.63$ & $ -1.14$\\
$ 5$ & $ 0.35$ & $ 0.00$ & $ -2.68$ & $  0.04$ & $ -6.14$ & $ -1.43$\\
$ 6$ & $ 0.35$ & $ 0.00$ & $ -3.06$ & $ -0.13$ & $ -6.56$ & $ -1.66$\\
$ 7$ & $ 0.35$ & $ 0.00$ & $ -3.28$ & $ -0.22$ & $ -6.78$ & $ -1.78$\\
$ 8$ & $ 0.35$ & $ 0.00$ & $ -3.50$ & $ -0.31$ & $ -7.00$ & $ -1.92$\\
$ 9$ & $ 0.35$ & $ 0.00$ & $ -3.70$ & $ -0.39$ & $ -7.19$ & $ -2.03$\\
$10$ & $ 0.35$ & $ 0.00$ & $ -3.95$ & $ -0.49$ & $ -7.43$ & $ -2.16$\\
$11$ & $ 0.35$ & $ 0.00$ & $ -4.21$ & $ -0.60$ & $ -7.67$ & $ -2.30$\\
$12$ & $ 0.35$ & $ 0.00$ & $ -4.42$ & $ -0.69$ & $ -7.86$ & $ -2.41$\\
$13$ & $ 0.35$ & $ 0.00$ & $ -4.60$ & $ -0.76$ & $ -8.03$ & $ -2.50$\\
$14$ & $ 0.35$ & $ 0.00$ & $ -4.77$ & $ -0.84$ & $ -8.16$ & $ -2.58$\\
$15$ & $ 0.35$ & $ 0.00$ & $ -4.94$ & $ -0.90$ & $ -8.30$ & $ -2.66$\\
$ 1$ & $ 0.67$ & $ 0.00$ & $  1.83$ & $  2.14$ & $ -0.41$ & $  1.60$\\
$ 2$ & $ 0.67$ & $ 0.00$ & $ -1.80$ & $  0.57$ & $ -4.85$ & $ -0.62$\\
$ 3$ & $ 0.67$ & $ 0.00$ & $ -3.34$ & $ -0.10$ & $ -6.67$ & $ -1.62$\\
$ 4$ & $ 0.67$ & $ 0.00$ & $ -4.18$ & $ -0.47$ & $ -7.58$ & $ -2.11$\\
$ 5$ & $ 0.67$ & $ 0.00$ & $ -4.44$ & $ -0.59$ & $ -7.84$ & $ -2.26$\\
$ 6$ & $ 0.67$ & $ 0.00$ & $ -4.62$ & $ -0.67$ & $ -8.02$ & $ -2.36$\\
$ 7$ & $ 0.67$ & $ 0.00$ & $ -4.83$ & $ -0.76$ & $ -8.22$ & $ -2.48$\\
$ 8$ & $ 0.67$ & $ 0.00$ & $ -5.06$ & $ -0.86$ & $ -8.44$ & $ -2.61$\\
$ 9$ & $ 0.67$ & $ 0.00$ & $ -5.34$ & $ -0.98$ & $ -8.70$ & $ -2.75$\\
$10$ & $ 0.67$ & $ 0.00$ & $ -5.66$ & $ -1.11$ & $ -8.99$ & $ -2.92$\\
$11$ & $ 0.67$ & $ 0.00$ & $ -5.79$ & $ -1.18$ & $ -9.10$ & $ -2.99$\\
$12$ & $ 0.67$ & $ 0.00$ & $ -5.93$ & $ -1.25$ & $ -9.22$ & $ -3.06$\\
$13$ & $ 0.67$ & $ 0.00$ & $ -6.06$ & $ -1.32$ & $ -9.33$ & $ -3.12$\\
$14$ & $ 0.67$ & $ 0.00$ & $ -6.15$ & $ -1.36$ & $ -9.40$ & $ -3.16$\\
$15$ & $ 0.67$ & $ 0.00$ & $ -6.22$ & $ -1.40$ & $ -9.46$ & $ -3.20$\\
\hline
\end{tabular}
\end{minipage}
\end{table*}

\clearpage
\begin{table*}
\caption{Simple stellar population models for the higher-order Balmer
lines with $t=1\;$--$\;15\;$Gyr, $[\ZH]=-2.25\;$--$\;0.67$, and
$[\aFe]=0.3$.}
\begin{minipage}{0.49\textwidth}
\begin{tabular}{rrrrrrr}
\hline\hline
\multicolumn{1}{c}{Age} & \multicolumn{1}{c}{[\ZH]} & \multicolumn{1}{c}{[\aFe]} & \multicolumn{1}{c}{\HdA} & \multicolumn{1}{c}{\HdF} & \multicolumn{1}{c}{\HgA} & \multicolumn{1}{c}{\HgF}\\
\multicolumn{1}{c}{(1)} & \multicolumn{1}{c}{(2)}   & \multicolumn{1}{c}{(3)}    & \multicolumn{1}{c}{(4)}  & \multicolumn{1}{c}{(5)}  & \multicolumn{1}{c}{(6)} & \multicolumn{1}{c}{(7)}\\
\hline
$ 1$ & $-2.25$ & $ 0.30$ & $ 10.65$ & $  8.30$ & $ 10.47$ & $  8.09$\\
$ 2$ & $-2.25$ & $ 0.30$ & $  8.50$ & $  6.33$ & $  7.70$ & $  6.05$\\
$ 3$ & $-2.25$ & $ 0.30$ & $  8.18$ & $  5.94$ & $  6.88$ & $  5.48$\\
$ 4$ & $-2.25$ & $ 0.30$ & $  7.77$ & $  5.58$ & $  6.24$ & $  5.07$\\
$ 5$ & $-2.25$ & $ 0.30$ & $  7.24$ & $  5.19$ & $  5.54$ & $  4.63$\\
$ 6$ & $-2.25$ & $ 0.30$ & $  6.73$ & $  4.82$ & $  4.86$ & $  4.22$\\
$ 7$ & $-2.25$ & $ 0.30$ & $  6.19$ & $  4.49$ & $  4.29$ & $  3.88$\\
$ 8$ & $-2.25$ & $ 0.30$ & $  5.84$ & $  4.29$ & $  3.99$ & $  3.68$\\
$ 9$ & $-2.25$ & $ 0.30$ & $  5.54$ & $  4.25$ & $  3.33$ & $  3.43$\\
$10$ & $-2.25$ & $ 0.30$ & $  5.71$ & $  4.30$ & $  3.32$ & $  3.41$\\
$11$ & $-2.25$ & $ 0.30$ & $  5.91$ & $  4.44$ & $  3.62$ & $  3.54$\\
$12$ & $-2.25$ & $ 0.30$ & $  5.97$ & $  4.53$ & $  3.74$ & $  3.60$\\
$13$ & $-2.25$ & $ 0.30$ & $  5.76$ & $  4.41$ & $  3.59$ & $  3.46$\\
$14$ & $-2.25$ & $ 0.30$ & $  5.34$ & $  4.15$ & $  3.05$ & $  3.12$\\
$15$ & $-2.25$ & $ 0.30$ & $  5.11$ & $  4.00$ & $  2.84$ & $  2.96$\\
$ 1$ & $-1.35$ & $ 0.30$ & $  7.98$ & $  5.92$ & $  7.07$ & $  5.68$\\
$ 2$ & $-1.35$ & $ 0.30$ & $  7.60$ & $  5.49$ & $  6.49$ & $  5.24$\\
$ 3$ & $-1.35$ & $ 0.30$ & $  6.88$ & $  4.98$ & $  5.46$ & $  4.65$\\
$ 4$ & $-1.35$ & $ 0.30$ & $  5.97$ & $  4.39$ & $  4.25$ & $  3.97$\\
$ 5$ & $-1.35$ & $ 0.30$ & $  5.08$ & $  3.83$ & $  3.05$ & $  3.30$\\
$ 6$ & $-1.35$ & $ 0.30$ & $  4.39$ & $  3.41$ & $  2.17$ & $  2.80$\\
$ 7$ & $-1.35$ & $ 0.30$ & $  3.84$ & $  3.06$ & $  1.46$ & $  2.38$\\
$ 8$ & $-1.35$ & $ 0.30$ & $  3.51$ & $  2.86$ & $  1.07$ & $  2.15$\\
$ 9$ & $-1.35$ & $ 0.30$ & $  3.17$ & $  2.66$ & $  0.56$ & $  1.86$\\
$10$ & $-1.35$ & $ 0.30$ & $  3.00$ & $  2.52$ & $  0.32$ & $  1.70$\\
$11$ & $-1.35$ & $ 0.30$ & $  2.94$ & $  2.47$ & $  0.26$ & $  1.64$\\
$12$ & $-1.35$ & $ 0.30$ & $  3.05$ & $  2.44$ & $  0.43$ & $  1.65$\\
$13$ & $-1.35$ & $ 0.30$ & $  3.32$ & $  2.67$ & $  0.73$ & $  1.85$\\
$14$ & $-1.35$ & $ 0.30$ & $  4.02$ & $  3.14$ & $  1.52$ & $  2.29$\\
$15$ & $-1.35$ & $ 0.30$ & $  4.34$ & $  3.34$ & $  1.86$ & $  2.46$\\
$ 1$ & $-0.33$ & $ 0.30$ & $  6.14$ & $  4.37$ & $  4.32$ & $  3.95$\\
$ 2$ & $-0.33$ & $ 0.30$ & $  3.33$ & $  2.82$ & $  0.77$ & $  2.09$\\
$ 3$ & $-0.33$ & $ 0.30$ & $  2.21$ & $  2.21$ & $ -0.62$ & $  1.34$\\
$ 4$ & $-0.33$ & $ 0.30$ & $  1.56$ & $  1.82$ & $ -1.45$ & $  0.86$\\
$ 5$ & $-0.33$ & $ 0.30$ & $  1.18$ & $  1.60$ & $ -2.01$ & $  0.55$\\
$ 6$ & $-0.33$ & $ 0.30$ & $  0.87$ & $  1.43$ & $ -2.46$ & $  0.30$\\
$ 7$ & $-0.33$ & $ 0.30$ & $  0.59$ & $  1.28$ & $ -2.85$ & $  0.07$\\
$ 8$ & $-0.33$ & $ 0.30$ & $  0.37$ & $  1.16$ & $ -3.16$ & $ -0.10$\\
$ 9$ & $-0.33$ & $ 0.30$ & $  0.15$ & $  1.05$ & $ -3.46$ & $ -0.27$\\
$10$ & $-0.33$ & $ 0.30$ & $ -0.12$ & $  0.91$ & $ -3.80$ & $ -0.47$\\
$11$ & $-0.33$ & $ 0.30$ & $ -0.33$ & $  0.80$ & $ -4.07$ & $ -0.62$\\
$12$ & $-0.33$ & $ 0.30$ & $ -0.49$ & $  0.71$ & $ -4.27$ & $ -0.75$\\
$13$ & $-0.33$ & $ 0.30$ & $ -0.62$ & $  0.66$ & $ -4.43$ & $ -0.83$\\
$14$ & $-0.33$ & $ 0.30$ & $ -0.65$ & $  0.66$ & $ -4.48$ & $ -0.86$\\
$15$ & $-0.33$ & $ 0.30$ & $ -0.74$ & $  0.63$ & $ -4.59$ & $ -0.93$\\
\hline
\end{tabular}
\end{minipage}
\begin{minipage}{0.49\textwidth}
\begin{tabular}{rrrrrrr}
\hline\hline
\multicolumn{1}{c}{Age} & \multicolumn{1}{c}{[\ZH]} & \multicolumn{1}{c}{[\aFe]} & \multicolumn{1}{c}{\HdA} & \multicolumn{1}{c}{\HdF} & \multicolumn{1}{c}{\HgA} & \multicolumn{1}{c}{\HgF}\\
\multicolumn{1}{c}{(1)} & \multicolumn{1}{c}{(2)}   & \multicolumn{1}{c}{(3)}    & \multicolumn{1}{c}{(4)}  & \multicolumn{1}{c}{(5)}  & \multicolumn{1}{c}{(6)} & \multicolumn{1}{c}{(7)}\\
\hline
$ 1$ & $ 0.00$ & $ 0.30$ & $  6.22$ & $  4.32$ & $  4.40$ & $  3.91$\\
$ 2$ & $ 0.00$ & $ 0.30$ & $  1.94$ & $  2.10$ & $ -0.85$ & $  1.17$\\
$ 3$ & $ 0.00$ & $ 0.30$ & $  0.71$ & $  1.46$ & $ -2.52$ & $  0.27$\\
$ 4$ & $ 0.00$ & $ 0.30$ & $  0.23$ & $  1.21$ & $ -3.19$ & $ -0.10$\\
$ 5$ & $ 0.00$ & $ 0.30$ & $ -0.22$ & $  0.99$ & $ -3.79$ & $ -0.43$\\
$ 6$ & $ 0.00$ & $ 0.30$ & $ -0.58$ & $  0.81$ & $ -4.24$ & $ -0.69$\\
$ 7$ & $ 0.00$ & $ 0.30$ & $ -0.79$ & $  0.72$ & $ -4.48$ & $ -0.83$\\
$ 8$ & $ 0.00$ & $ 0.30$ & $ -0.92$ & $  0.65$ & $ -4.65$ & $ -0.94$\\
$ 9$ & $ 0.00$ & $ 0.30$ & $ -1.16$ & $  0.55$ & $ -4.92$ & $ -1.09$\\
$10$ & $ 0.00$ & $ 0.30$ & $ -1.32$ & $  0.48$ & $ -5.11$ & $ -1.21$\\
$11$ & $ 0.00$ & $ 0.30$ & $ -1.55$ & $  0.38$ & $ -5.36$ & $ -1.35$\\
$12$ & $ 0.00$ & $ 0.30$ & $ -1.73$ & $  0.30$ & $ -5.55$ & $ -1.47$\\
$13$ & $ 0.00$ & $ 0.30$ & $ -1.89$ & $  0.23$ & $ -5.72$ & $ -1.57$\\
$14$ & $ 0.00$ & $ 0.30$ & $ -2.03$ & $  0.17$ & $ -5.86$ & $ -1.66$\\
$15$ & $ 0.00$ & $ 0.30$ & $ -2.17$ & $  0.11$ & $ -5.99$ & $ -1.75$\\
$ 1$ & $ 0.35$ & $ 0.30$ & $  3.93$ & $  3.04$ & $  1.19$ & $  2.17$\\
$ 2$ & $ 0.35$ & $ 0.30$ & $  0.42$ & $  1.48$ & $ -2.90$ & $  0.12$\\
$ 3$ & $ 0.35$ & $ 0.30$ & $ -0.27$ & $  1.12$ & $ -3.83$ & $ -0.39$\\
$ 4$ & $ 0.35$ & $ 0.30$ & $ -0.84$ & $  0.85$ & $ -4.54$ & $ -0.79$\\
$ 5$ & $ 0.35$ & $ 0.30$ & $ -1.24$ & $  0.67$ & $ -5.01$ & $ -1.07$\\
$ 6$ & $ 0.35$ & $ 0.30$ & $ -1.57$ & $  0.52$ & $ -5.39$ & $ -1.29$\\
$ 7$ & $ 0.35$ & $ 0.30$ & $ -1.76$ & $  0.44$ & $ -5.60$ & $ -1.42$\\
$ 8$ & $ 0.35$ & $ 0.30$ & $ -1.95$ & $  0.35$ & $ -5.80$ & $ -1.55$\\
$ 9$ & $ 0.35$ & $ 0.30$ & $ -2.12$ & $  0.28$ & $ -5.97$ & $ -1.66$\\
$10$ & $ 0.35$ & $ 0.30$ & $ -2.31$ & $  0.20$ & $ -6.17$ & $ -1.78$\\
$11$ & $ 0.35$ & $ 0.30$ & $ -2.51$ & $  0.11$ & $ -6.37$ & $ -1.91$\\
$12$ & $ 0.35$ & $ 0.30$ & $ -2.67$ & $  0.04$ & $ -6.52$ & $ -2.01$\\
$13$ & $ 0.35$ & $ 0.30$ & $ -2.80$ & $ -0.01$ & $ -6.65$ & $ -2.09$\\
$14$ & $ 0.35$ & $ 0.30$ & $ -2.91$ & $ -0.06$ & $ -6.75$ & $ -2.17$\\
$15$ & $ 0.35$ & $ 0.30$ & $ -3.01$ & $ -0.11$ & $ -6.84$ & $ -2.23$\\
$ 1$ & $ 0.67$ & $ 0.30$ & $  3.03$ & $  2.66$ & $  0.33$ & $  1.83$\\
$ 2$ & $ 0.67$ & $ 0.30$ & $ -0.41$ & $  1.17$ & $ -3.96$ & $ -0.35$\\
$ 3$ & $ 0.67$ & $ 0.30$ & $ -1.77$ & $  0.56$ & $ -5.64$ & $ -1.31$\\
$ 4$ & $ 0.67$ & $ 0.30$ & $ -2.47$ & $  0.24$ & $ -6.43$ & $ -1.78$\\
$ 5$ & $ 0.67$ & $ 0.30$ & $ -2.69$ & $  0.13$ & $ -6.68$ & $ -1.92$\\
$ 6$ & $ 0.67$ & $ 0.30$ & $ -2.83$ & $  0.07$ & $ -6.83$ & $ -2.02$\\
$ 7$ & $ 0.67$ & $ 0.30$ & $ -2.99$ & $  0.00$ & $ -6.99$ & $ -2.14$\\
$ 8$ & $ 0.67$ & $ 0.30$ & $ -3.17$ & $ -0.09$ & $ -7.18$ & $ -2.26$\\
$ 9$ & $ 0.67$ & $ 0.30$ & $ -3.37$ & $ -0.18$ & $ -7.39$ & $ -2.40$\\
$10$ & $ 0.67$ & $ 0.30$ & $ -3.62$ & $ -0.29$ & $ -7.63$ & $ -2.56$\\
$11$ & $ 0.67$ & $ 0.30$ & $ -3.69$ & $ -0.34$ & $ -7.71$ & $ -2.62$\\
$12$ & $ 0.67$ & $ 0.30$ & $ -3.78$ & $ -0.40$ & $ -7.79$ & $ -2.68$\\
$13$ & $ 0.67$ & $ 0.30$ & $ -3.84$ & $ -0.44$ & $ -7.86$ & $ -2.74$\\
$14$ & $ 0.67$ & $ 0.30$ & $ -3.87$ & $ -0.47$ & $ -7.89$ & $ -2.78$\\
$15$ & $ 0.67$ & $ 0.30$ & $ -3.89$ & $ -0.49$ & $ -7.91$ & $ -2.81$\\
\hline
\end{tabular}
\end{minipage}
\end{table*}

\clearpage
\begin{table*}
\caption{Simple stellar population models for the higher-order Balmer
lines with $t=1\;$--$\;15\;$Gyr, $[\ZH]=-2.25\;$--$\;0.67$, and
$[\aFe]=0.5$.}
\begin{minipage}{0.49\textwidth}
\begin{tabular}{rrrrrrr}
\hline\hline
\multicolumn{1}{c}{Age} & \multicolumn{1}{c}{[\ZH]} & \multicolumn{1}{c}{[\aFe]} & \multicolumn{1}{c}{\HdA} & \multicolumn{1}{c}{\HdF} & \multicolumn{1}{c}{\HgA} & \multicolumn{1}{c}{\HgF}\\
\multicolumn{1}{c}{(1)} & \multicolumn{1}{c}{(2)}   & \multicolumn{1}{c}{(3)}    & \multicolumn{1}{c}{(4)}  & \multicolumn{1}{c}{(5)}  & \multicolumn{1}{c}{(6)} & \multicolumn{1}{c}{(7)}\\
\hline
$ 1$ & $-2.25$ & $ 0.50$ & $ 10.74$ & $  8.33$ & $ 10.46$ & $  8.08$\\
$ 2$ & $-2.25$ & $ 0.50$ & $  8.57$ & $  6.36$ & $  7.70$ & $  6.04$\\
$ 3$ & $-2.25$ & $ 0.50$ & $  8.26$ & $  5.97$ & $  6.87$ & $  5.47$\\
$ 4$ & $-2.25$ & $ 0.50$ & $  7.85$ & $  5.61$ & $  6.24$ & $  5.06$\\
$ 5$ & $-2.25$ & $ 0.50$ & $  7.33$ & $  5.22$ & $  5.54$ & $  4.62$\\
$ 6$ & $-2.25$ & $ 0.50$ & $  6.82$ & $  4.85$ & $  4.85$ & $  4.21$\\
$ 7$ & $-2.25$ & $ 0.50$ & $  6.28$ & $  4.53$ & $  4.28$ & $  3.87$\\
$ 8$ & $-2.25$ & $ 0.50$ & $  5.94$ & $  4.33$ & $  3.98$ & $  3.66$\\
$ 9$ & $-2.25$ & $ 0.50$ & $  5.64$ & $  4.29$ & $  3.33$ & $  3.42$\\
$10$ & $-2.25$ & $ 0.50$ & $  5.81$ & $  4.34$ & $  3.31$ & $  3.39$\\
$11$ & $-2.25$ & $ 0.50$ & $  6.01$ & $  4.48$ & $  3.61$ & $  3.53$\\
$12$ & $-2.25$ & $ 0.50$ & $  6.07$ & $  4.57$ & $  3.73$ & $  3.59$\\
$13$ & $-2.25$ & $ 0.50$ & $  5.86$ & $  4.45$ & $  3.58$ & $  3.44$\\
$14$ & $-2.25$ & $ 0.50$ & $  5.44$ & $  4.19$ & $  3.04$ & $  3.11$\\
$15$ & $-2.25$ & $ 0.50$ & $  5.21$ & $  4.04$ & $  2.83$ & $  2.94$\\
$ 1$ & $-1.35$ & $ 0.50$ & $  8.08$ & $  5.97$ & $  7.13$ & $  5.68$\\
$ 2$ & $-1.35$ & $ 0.50$ & $  7.70$ & $  5.55$ & $  6.56$ & $  5.25$\\
$ 3$ & $-1.35$ & $ 0.50$ & $  6.99$ & $  5.04$ & $  5.53$ & $  4.65$\\
$ 4$ & $-1.35$ & $ 0.50$ & $  6.09$ & $  4.45$ & $  4.33$ & $  3.98$\\
$ 5$ & $-1.35$ & $ 0.50$ & $  5.21$ & $  3.89$ & $  3.13$ & $  3.30$\\
$ 6$ & $-1.35$ & $ 0.50$ & $  4.53$ & $  3.48$ & $  2.26$ & $  2.80$\\
$ 7$ & $-1.35$ & $ 0.50$ & $  3.98$ & $  3.13$ & $  1.55$ & $  2.39$\\
$ 8$ & $-1.35$ & $ 0.50$ & $  3.66$ & $  2.94$ & $  1.16$ & $  2.16$\\
$ 9$ & $-1.35$ & $ 0.50$ & $  3.32$ & $  2.74$ & $  0.66$ & $  1.87$\\
$10$ & $-1.35$ & $ 0.50$ & $  3.16$ & $  2.60$ & $  0.42$ & $  1.71$\\
$11$ & $-1.35$ & $ 0.50$ & $  3.11$ & $  2.55$ & $  0.36$ & $  1.65$\\
$12$ & $-1.35$ & $ 0.50$ & $  3.22$ & $  2.53$ & $  0.54$ & $  1.66$\\
$13$ & $-1.35$ & $ 0.50$ & $  3.49$ & $  2.76$ & $  0.84$ & $  1.86$\\
$14$ & $-1.35$ & $ 0.50$ & $  4.19$ & $  3.22$ & $  1.63$ & $  2.30$\\
$15$ & $-1.35$ & $ 0.50$ & $  4.51$ & $  3.43$ & $  1.97$ & $  2.47$\\
$ 1$ & $-0.33$ & $ 0.50$ & $  6.55$ & $  4.55$ & $  4.55$ & $  4.02$\\
$ 2$ & $-0.33$ & $ 0.50$ & $  3.80$ & $  3.01$ & $  1.05$ & $  2.17$\\
$ 3$ & $-0.33$ & $ 0.50$ & $  2.69$ & $  2.41$ & $ -0.33$ & $  1.42$\\
$ 4$ & $-0.33$ & $ 0.50$ & $  2.06$ & $  2.03$ & $ -1.15$ & $  0.95$\\
$ 5$ & $-0.33$ & $ 0.50$ & $  1.69$ & $  1.82$ & $ -1.69$ & $  0.64$\\
$ 6$ & $-0.33$ & $ 0.50$ & $  1.39$ & $  1.65$ & $ -2.14$ & $  0.39$\\
$ 7$ & $-0.33$ & $ 0.50$ & $  1.12$ & $  1.50$ & $ -2.52$ & $  0.17$\\
$ 8$ & $-0.33$ & $ 0.50$ & $  0.91$ & $  1.39$ & $ -2.82$ & $  0.00$\\
$ 9$ & $-0.33$ & $ 0.50$ & $  0.70$ & $  1.28$ & $ -3.11$ & $ -0.18$\\
$10$ & $-0.33$ & $ 0.50$ & $  0.45$ & $  1.14$ & $ -3.43$ & $ -0.37$\\
$11$ & $-0.33$ & $ 0.50$ & $  0.25$ & $  1.04$ & $ -3.70$ & $ -0.52$\\
$12$ & $-0.33$ & $ 0.50$ & $  0.11$ & $  0.96$ & $ -3.88$ & $ -0.64$\\
$13$ & $-0.33$ & $ 0.50$ & $  0.00$ & $  0.92$ & $ -4.03$ & $ -0.72$\\
$14$ & $-0.33$ & $ 0.50$ & $ -0.01$ & $  0.92$ & $ -4.06$ & $ -0.74$\\
$15$ & $-0.33$ & $ 0.50$ & $ -0.09$ & $  0.89$ & $ -4.16$ & $ -0.80$\\
\hline
\end{tabular}
\end{minipage}
\begin{minipage}{0.49\textwidth}
\begin{tabular}{rrrrrrr}
\hline\hline
\multicolumn{1}{c}{Age} & \multicolumn{1}{c}{[\ZH]} & \multicolumn{1}{c}{[\aFe]} & \multicolumn{1}{c}{\HdA} & \multicolumn{1}{c}{\HdF} & \multicolumn{1}{c}{\HgA} & \multicolumn{1}{c}{\HgF}\\
\multicolumn{1}{c}{(1)} & \multicolumn{1}{c}{(2)}   & \multicolumn{1}{c}{(3)}    & \multicolumn{1}{c}{(4)}  & \multicolumn{1}{c}{(5)}  & \multicolumn{1}{c}{(6)} & \multicolumn{1}{c}{(7)}\\
\hline
$ 1$ & $ 0.00$ & $ 0.50$ & $  6.66$ & $  4.52$ & $  4.74$ & $  4.02$\\
$ 2$ & $ 0.00$ & $ 0.50$ & $  2.50$ & $  2.35$ & $ -0.41$ & $  1.31$\\
$ 3$ & $ 0.00$ & $ 0.50$ & $  1.37$ & $  1.75$ & $ -1.97$ & $  0.44$\\
$ 4$ & $ 0.00$ & $ 0.50$ & $  0.90$ & $  1.50$ & $ -2.63$ & $  0.08$\\
$ 5$ & $ 0.00$ & $ 0.50$ & $  0.49$ & $  1.29$ & $ -3.20$ & $ -0.25$\\
$ 6$ & $ 0.00$ & $ 0.50$ & $  0.14$ & $  1.12$ & $ -3.64$ & $ -0.50$\\
$ 7$ & $ 0.00$ & $ 0.50$ & $ -0.05$ & $  1.03$ & $ -3.87$ & $ -0.64$\\
$ 8$ & $ 0.00$ & $ 0.50$ & $ -0.19$ & $  0.97$ & $ -4.04$ & $ -0.75$\\
$ 9$ & $ 0.00$ & $ 0.50$ & $ -0.40$ & $  0.87$ & $ -4.29$ & $ -0.90$\\
$10$ & $ 0.00$ & $ 0.50$ & $ -0.55$ & $  0.80$ & $ -4.47$ & $ -1.01$\\
$11$ & $ 0.00$ & $ 0.50$ & $ -0.75$ & $  0.71$ & $ -4.70$ & $ -1.15$\\
$12$ & $ 0.00$ & $ 0.50$ & $ -0.90$ & $  0.65$ & $ -4.87$ & $ -1.26$\\
$13$ & $ 0.00$ & $ 0.50$ & $ -1.03$ & $  0.59$ & $ -5.02$ & $ -1.36$\\
$14$ & $ 0.00$ & $ 0.50$ & $ -1.14$ & $  0.53$ & $ -5.14$ & $ -1.44$\\
$15$ & $ 0.00$ & $ 0.50$ & $ -1.25$ & $  0.48$ & $ -5.25$ & $ -1.52$\\
$ 1$ & $ 0.35$ & $ 0.50$ & $  4.62$ & $  3.35$ & $  1.71$ & $  2.34$\\
$ 2$ & $ 0.35$ & $ 0.50$ & $  1.28$ & $  1.87$ & $ -2.23$ & $  0.33$\\
$ 3$ & $ 0.35$ & $ 0.50$ & $  0.60$ & $  1.51$ & $ -3.14$ & $ -0.17$\\
$ 4$ & $ 0.35$ & $ 0.50$ & $  0.07$ & $  1.26$ & $ -3.82$ & $ -0.57$\\
$ 5$ & $ 0.35$ & $ 0.50$ & $ -0.31$ & $  1.08$ & $ -4.28$ & $ -0.84$\\
$ 6$ & $ 0.35$ & $ 0.50$ & $ -0.60$ & $  0.94$ & $ -4.63$ & $ -1.05$\\
$ 7$ & $ 0.35$ & $ 0.50$ & $ -0.78$ & $  0.86$ & $ -4.83$ & $ -1.18$\\
$ 8$ & $ 0.35$ & $ 0.50$ & $ -0.95$ & $  0.79$ & $ -5.02$ & $ -1.31$\\
$ 9$ & $ 0.35$ & $ 0.50$ & $ -1.10$ & $  0.72$ & $ -5.18$ & $ -1.41$\\
$10$ & $ 0.35$ & $ 0.50$ & $ -1.26$ & $  0.65$ & $ -5.35$ & $ -1.54$\\
$11$ & $ 0.35$ & $ 0.50$ & $ -1.42$ & $  0.58$ & $ -5.53$ & $ -1.66$\\
$12$ & $ 0.35$ & $ 0.50$ & $ -1.55$ & $  0.52$ & $ -5.65$ & $ -1.75$\\
$13$ & $ 0.35$ & $ 0.50$ & $ -1.64$ & $  0.48$ & $ -5.76$ & $ -1.83$\\
$14$ & $ 0.35$ & $ 0.50$ & $ -1.71$ & $  0.44$ & $ -5.83$ & $ -1.90$\\
$15$ & $ 0.35$ & $ 0.50$ & $ -1.78$ & $  0.41$ & $ -5.89$ & $ -1.96$\\
$ 1$ & $ 0.67$ & $ 0.50$ & $  3.82$ & $  3.02$ & $  0.81$ & $  1.98$\\
$ 2$ & $ 0.67$ & $ 0.50$ & $  0.50$ & $  1.56$ & $ -3.37$ & $ -0.17$\\
$ 3$ & $ 0.67$ & $ 0.50$ & $ -0.74$ & $  1.00$ & $ -4.96$ & $ -1.11$\\
$ 4$ & $ 0.67$ & $ 0.50$ & $ -1.36$ & $  0.71$ & $ -5.68$ & $ -1.57$\\
$ 5$ & $ 0.67$ & $ 0.50$ & $ -1.56$ & $  0.61$ & $ -5.92$ & $ -1.71$\\
$ 6$ & $ 0.67$ & $ 0.50$ & $ -1.67$ & $  0.55$ & $ -6.05$ & $ -1.81$\\
$ 7$ & $ 0.67$ & $ 0.50$ & $ -1.80$ & $  0.49$ & $ -6.19$ & $ -1.91$\\
$ 8$ & $ 0.67$ & $ 0.50$ & $ -1.94$ & $  0.42$ & $ -6.35$ & $ -2.03$\\
$ 9$ & $ 0.67$ & $ 0.50$ & $ -2.11$ & $  0.34$ & $ -6.53$ & $ -2.17$\\
$10$ & $ 0.67$ & $ 0.50$ & $ -2.30$ & $  0.24$ & $ -6.74$ & $ -2.32$\\
$11$ & $ 0.67$ & $ 0.50$ & $ -2.35$ & $  0.20$ & $ -6.80$ & $ -2.38$\\
$12$ & $ 0.67$ & $ 0.50$ & $ -2.40$ & $  0.16$ & $ -6.86$ & $ -2.44$\\
$13$ & $ 0.67$ & $ 0.50$ & $ -2.42$ & $  0.13$ & $ -6.90$ & $ -2.49$\\
$14$ & $ 0.67$ & $ 0.50$ & $ -2.41$ & $  0.11$ & $ -6.91$ & $ -2.53$\\
$15$ & $ 0.67$ & $ 0.50$ & $ -2.39$ & $  0.11$ & $ -6.90$ & $ -2.55$\\
\hline
\end{tabular}
\end{minipage}
\end{table*}

\end{document}